\documentstyle{article}
\newcommand{\K}{\mbox{\bf K}}
\newcommand{\R}{\mbox{\bf R}}
\newcommand{\C}{\mbox{\bf C}}
\newcommand{\Om}{{\bf\Omega}}
\newcommand{\e}{\mbox{\bf e}}
\newcommand{\p}{\prime}
\newcommand{\M}{\mbox{M}}
\newcommand{\s}{\scriptstyle}
\newtheorem{theorem}{Theorem}
\begin{document}
\begin{titlepage}
\title{\huge Modulo 2 periodicity of complex Clifford algebras and 
electromagnetic field}
\author{Vadim V. Varlamov\\
{\small\it Applied Mathematics, Siberian State Academy of Mining}\hspace{1mm}
\small{\&}\hspace{1mm}{\small\it Metallurgy,}\\
{\small\it Novokuznetsk, Russia\thanks{E-mail: root@varlamov.kemerovo.su}}}
\date{}
\maketitle
\vspace{2cm}
\begin{abstract}Electromagnetic field is considered in the framework of a
Clifford algebra $\C_{2}$ over a field of complex numbers. It is shown here
that a modulo 2 periodicity of complex Clifford algebras may be connected
with electromagnetic field. \end{abstract}
\thispagestyle{empty}
\end{titlepage}
\newpage
\section{INTRODUCTION}
It is well-known that algebra introduced by Clifford \cite{1} be
simultaneously a synthesis and generalization of Hamilton theory of quaternions
and Grassmann algebras. In the other words, Clifford algebra is generalization
of quaternion algebra onto a case of manydimensional spaces. Thanks to this
fact Lipschitz \cite{2} was determined the tight connection between
Clifford algebras and rotation groups of manydimensional spaces. The Clifford
algebras becomes the object of fixed attention of physicists after introduction
in theory of electron the noncommutative algebra of Dirac $\gamma$-matricies.

We shall denote further Clifford algebra as ${}^{l}\K_{n}$, where $\K$ is
algebraic field of characteristic 0 $(\K=\R,\K=\Om,\K=\C)$; $\Om$ is a field
of double numbers (so-called singular field \cite{3}).

In essence, the problem of periodicity of algebras ${}^{l}\K_{n}$ to begin on
the first Clifford paper (see ref.1), where the each algebra ${}^{l}\K_{n}$ is
considered as a product of the some number of quaternion algebras. After
construction of spinor representation of Clifford algebras \cite{4}
the structure of periodicity was finally elucidated by Atiyah, Bott, and
Shapiro \cite{5}. Namely, the Clifford algebras over a field of real
numbers $\K=\R$ are possess a modulo 8 periodicity. By force of identity
$\Om=\R\oplus\R$ the algebras over a field of double numbers are possess the
same periodicity. In the case of a field $\K=\C$ we have a modulo 2 periodicity.
\begin{sloppypar}
The application of these periodicities in physics to begin on the paper of
Coquereaux \cite{6}, where the modulo 8 periodicity of real algebras
was used to particle physics. In present paper the modulo 2 periodicity of
complex algebras is connected with electromagnetic field. This connection is
possible by force of realization of the basic notions of electromagnetic
field (such that vectors {\bf E} and {\bf H}, Maxwell equations) in the terms
of an algebra $\C_{2}$, which by Clifford terminology is called the algebra of
hyperbolic biquaternions.\end{sloppypar}
\section{CLIFFORD ALGEBRAS OVER A FIELD $\K$}
A Clifford algebra $\R_{n-1}$ over a field of real numbers $\R$ be an algebra
with $2^{n-1}$ basic elements: $\e_{0}$ (unit of algebra), $\e_{1},\e_{2},
\ldots,\e_{n-1}$ and products $\e_{i_{1}i_{2}\ldots i_{k}}=\e_{i_{1}}
\e_{i_{2}}\cdots \e_{i_{k}}$. Multiplication in $\R_{n-1}$ defined by 
a following rule:
\begin{equation}\label{e1}
\e^{2}_{i}=\e_{0}\hspace{2mm},\hspace{2mm} \e_{i}\e_{j}=-\e_{j}\e_{i}
\end{equation}

Furtner on, ${}^{l}\R_{n-1}$ be an algebra which we obtain from $\R_{n-1}$ by
substitution (\ref{e1}) onto
\begin{equation}\label{e2}
\e^{2}_{i}=\sigma(i-l)\e_{0}\hspace{2mm},\hspace{2mm} \e_{i}\e_{j}=
-\e_{j}\e_{i},
\end{equation}
where
$$
\sigma(n)=\left\{
\begin{array}{rl}
-1, & \mbox{if}\; n\le 0 \\
 1, & \mbox{if}\; n > 0  \\
\end{array} \right.
$$

The algebra $\R_{n-1}$ be a particular case of ${}^{l}\R_{n-1}$ when $l=0$.

A general element ${\cal A}$ of ${}^{l}\R_{n-1}$ represented by a following
formal polynomial:
$${\cal A}=a_{0}\e_{0} + \sum_{i=1}^{n-1}a_{i}\e_{i} + \sum_{i=1}^{n-1}
\sum_{j=1}^{n-1}a_{ij}\e_{ij}+\ldots +
\sum_{i_{1}=1}^{n-1}\cdots\sum_{i_{k}=1}^{n-1}a_{i_{1}\ldots i_{k}}\e_{i_{1}
\ldots i_{k}}+$$
$$+\ldots + a_{12\ldots n-1}\e_{12\ldots n-1}=\sum_{k=0}^{n-1}a_{i_{1}i_{2}
\ldots i_{k}}\e_{i_{1}i_{2}\ldots i_{k}}.$$

It is obvious that Clifford algebra ${}^{l}\R_{n-1}$ is associative. 
A center of
${}^{l}\R_{n-1}$ consist of unit $\e_{0}$ and volume element 
$\e_{12\ldots n-1}$.
The volume element is belong to the center only if $n-1$ is odd. Since
\begin{eqnarray}
\e_{12\ldots n-1}\e_{i}&=&\sigma(i-l)(-1)^{n-i-1}
\e_{12\ldots i-1 i+1\ldots n-1},\nonumber \\
\e_{i}\e_{12\ldots n-1}&=&\sigma(i-l)(-1)^{i-1}
\e_{12\ldots i-1 i+1\ldots n-1},\nonumber
\end{eqnarray}
then $\e_{12\ldots n-1}$ is belong to the center if 
$n-i-1\equiv i-1\!\!\!\!\pmod{2}$.
When $n-1$ is even, the element $\e_{12\ldots n-1}$ commutes with elements
$\e_{i_{1}i_{2}\ldots i_{2k+1}}$, and anticommutes with elements 
$\e_{i_{1}i_{2}\ldots i_{2k}}$.

The transition from ${}^{l}\R_{n-1}$ to ${}^{l}\R_{n}$ or 
${}^{l+1}\R_{n}$ may be
represented as the transition from the real numbers in ${}^{l}\R_{n-1}$ 
to complex coordinates $a+b\omega$, where $\omega$ is additional basis element 
$\e_{12\ldots n}$.
By force of (\ref{e1})-(\ref{e2}) in the case of transition from 
${}^{l}\R_{n-1}$ to ${}^{l}\R_{n}$
we have:
$$\e^{2}_{12\ldots n}=(-1)^{l+\frac{n(n-1)}{2}}$$
and in case ${}^{l}\R_{n-1}\rightarrow {}^{l+1}\R_{n}$:
$$\e_{12\ldots n}^{2}=(-1)^{l+1 +\frac{n(n-1)}{2}}.$$      

Therefore, in the first case
$$\omega^{2}=\left\{
\begin{array}{rl}
-1, & \mbox{if}\;n=4m+2\;\mbox{or}\;n=4m^{\p}-1,\;l\;\mbox{is even and}\\
    & \mbox{if}\;n=4m^{\p}\,\phantom{+2}\;\mbox{or}\;n=4m^{\p}+1,\;l\;
\mbox{is odd};\\
+1, & \mbox{if}\;n=4m^{\p}\,\phantom{+2}\;\mbox{or}\;n=4m^{\p}+1,\;l\;
\mbox{is even and}\\
    & \mbox{if}\;n=4m+2\;\mbox{or}\;n=4m^{\p}-1,\;l\;\mbox{is odd};
\end{array} \right. $$
and in the second case
$$\omega^{2}=\left\{
\begin{array}{rl}
-1, & \mbox{if}\;n=4m+2\;\mbox{or}\;n=4m^{\p}-1,\;l\;\mbox{is odd and}\\
    & \mbox{if}\;n=4m^{\p}\,\phantom{+2}\;\mbox{or}\;n=4m^{\p}+1,\;l\;
\mbox{is even};\\
+1, & \mbox{if}\;n=4m^{\p}\,\phantom{+2}\;\mbox{or}\;n=4m^{\p}+1,\;l\;
\mbox{is odd and}\\
    & \mbox{if}\;n=4m+2\;\mbox{or}\;n=4m^{\p}-1,\;l\;\mbox{is even};
\end{array} \right. $$
where $m=0,1,\ldots; m^{\p}=1,2,\ldots$.

Hence it follows that for $l$ is even we have:
\begin{equation}\label{e3}
{}^{l}\R_{4m^{\p}-1}={}^{l}\C_{4m^{\p}-2}\hspace{2mm},\hspace{2mm}{}^{l+1}
\R_{4m^{\p}+1}={}^{l}\C_{4m^{\p}},
\end{equation}
\begin{equation}\label{e4}
{}^{l}\R_{4m^{\p}+1}={}^{l}\Om_{4m^{\p}}\hspace{2mm},\hspace{2mm}{}^{l+1}
\R_{4m^{\p}-1}={}^{l}\Om_{4m^{\p}-2},
\end{equation}
and for $l$ is odd
\begin{equation}\label{e5}
{}^{l}\R_{4m^{\p}+1}={}^{l}\C_{4m^{\p}}\hspace{2mm},\hspace{2mm}{}^{l+1}
\R_{4m^{\p}-1}={}^{l}\C_{4m^{\p}-2},
\end{equation}
\begin{equation}\label{e6}
{}^{l}\R_{4m^{\p}-1}={}^{l}\Om_{4m^{\p}-2}\hspace{2mm},\hspace{2mm}{}^{l+1}
\R_{4m^{\p}+1}={}^{l}\Om_{4m^{\p}},
\end{equation}
where ${}^{l}\C_{n}$ and ${}^{l}\Om_{n}$ are Clifford algebras over the field
of complex numbers {\bf C} and field of double numbers $\Om$, respectively .
When $n$ is even, the identities of type (\ref{e3})-(\ref{e6}) excepted, 
since in this case
volume element $\omega=\e_{12\ldots n}$ is not belong to a center of extended
algebra.

In particular case, when $l=0$ we have a transition 
$\R_{n-1}\rightarrow \R_{n}$
and
$$\e^{2}_{12\ldots n}=(-1)^{\frac{n(n-1)}{2}}$$
Therefore,
$$\omega^{2}=\left\{
\begin{array}{rl}
-1, & \mbox{if}\;n=4m+2\;\mbox{or}\;n=4m^{\p}-1,\\
+1, & \mbox{if}\;n=4m^{\p}\,\phantom{+2}\;\mbox{or}\;n=4m^{\p}+1.
\end{array} \right. $$

In this case, for $n$ is odd we have the following identities:
\begin{equation}\label{e7}
\R_{4m^{\p}-1}=\C_{4m^{\p}-2}\hspace{2mm},\hspace{2mm}\R_{4m^{\p}+1}=
\Om_{4m^{\p}}.
\end{equation} 
\section{\hspace{-1mm}THE ALGEBRAS $\R_{3},\hspace{1mm}\C_{2}$ 
AND ELECTROMAGNETIC FIELD}
Consider now an algebra $\R_{3}$. The squares of units of this algebra 
are equal
to $\e^{2}_{i}=1 (i=0,1,2,3)$. The general element of $\R_{3}$ has a following
form:
$${\cal A}=a_{0}\e_{0}+\sum_{i=1}^{3}a_{i}\e_{i}+\sum_{i=1}^{3}
\sum_{j=1}^{3}a_{ij}\e_{ij}+a_{123}\e_{123}.$$

Further on, for a general element of $\R_{3}$ there exists a decomposition
${\cal A}^{3}={\cal A}^{2}+\omega{\cal A}^{2}$, where 
$\omega=\e_{123}\in \R_{3}$
and ${\cal A}^{2}$ is a general element of $\R_{2}$: 
$a_{0}\e_{0}+a_{1}\e_{1}+a_{2}\e_{2}+a_{12}\e_{12}$.
In fact
$${\cal A}^{3}={\cal A}^{2}+\omega{\cal A}^{2}=a_{0}\e_{0}+a_{1}\e_{1}+a_{2}
\e_{2}+a_{12}\e_{12}+\omega(b_{0}\e_{0}+b_{1}\e_{1}+b_{2}\e_{2}+b_{12}\e_{12})
=$$
\begin{equation}\label{e8}
(a_{0}+\omega b_{0})\e_{0}+(a_{1}+\omega b_{1})\e_{1}+(a_{2}+\omega b_{2})
\e_{2}+(a_{12}+\omega b_{12})\e_{12}=
\end{equation}
$$=a_{0}\e_{0}+a_{1}\e_{1}+a_{2}\e_{2}-b_{12}\e_{3}+a_{12}\e_{12}-b_{2}
\e_{13}+b_{1}\e_{23}+b_{0}\e_{123}.$$

Since in this case $\omega=\e_{123}$ is belong to the center of $\R_{3}$, and
$\omega^{2}=-1$, then in (\ref{e8}) the expressions $a_{k}+\omega b_{k}$ 
may be replaced
by $a_{k}+ib_{k}$, where $i$ is imaginary unit. This way, we have the
identity $\R_{3}=\C_{2}$, which be a particular case of (\ref{e7}). The algebra 
$\C_{2}$
is called the algebra of complex quaternions (or hyperbolic biquaternions).
Analogously, for the algebra ${}^{3}\R_{3}$ with $\omega^{2}=+1$ we have in
accordance with (\ref{e6}) the identity ${}^{3}\R_{3}={}^{2}\Om_{2}$, where
${}^{2}\Om_{2}$ is the algebra of elliptic biquaternions.

Further on, there exists a realization of electromagnetic field in the terms 
of $\R_{3}$. Let
\begin{equation}\label{e9}\left.
\begin{array}{lcr}
{\cal A}_{0}&=&\partial_{0}\e_{0}+\partial_{1}\e_{1}+\partial_{2}\e_{2}+
\partial_{3}\e_{3},\\
{\cal A}_{1}&=&A_{0}\e_{0}+A_{1}\e_{1}+A_{2}\e_{2}+A_{3}\e_{3},
\end{array}\right.
\end{equation}
where ${\cal A}_{0}$ and ${\cal A}_{1}$ are elements of $\R_{3}$. 
The coefficients
of these elements be partial derivatives and components of 
vector-potential, respectively.

Make up now the exterior product of elements (\ref{e9}):
$${\cal A}_{0}{\cal A}_{1}=(\partial_{0}\e_{0}+\partial_{1}\e_{1}+
\partial_{2}\e_{2}+\partial_{3}\e_{3})(A_{0}\e_{0}+A_{1}\e_{1}+A_{2}\e_{2}+
A_{3}\e_{3})=$$
$$=(\underbrace{\partial_{0}A_{0}+\partial_{1}A_{1}+\partial_{2}A_{2}+
\partial_{3}A_{3}}_{E_{0}})\e_{0}+(\underbrace{\partial_{0}A_{1}+
\partial_{1}A_{0}}_{E_{1}})\e_{0}\e_{1}+$$
\begin{equation}\label{e10}
(\underbrace{\partial_{0}A_{2}+\partial_{2}A_{0}}_{E_{2}})\e_{0}\e_{2}+
(\underbrace{\partial_{0}A_{3}+\partial_{3}A_{0}}_{E_{3}})\e_{0}\e_{3}+
(\underbrace{\partial_{2}A_{3}-\partial_{3}A_{2}}_{H_{1}})\e_{2}\e_{3}+
\end{equation}
$$+(\underbrace{\partial_{3}A_{1}-\partial_{1}A_{3}}_{H_{2}})\e_{3}\e_{1}+
(\underbrace{\partial_{1}A_{2}-\partial_{2}A_{1}}_{H_{3}})\e_{1}\e_{2}.$$

The scalar part $E_{0}\equiv 0$, since the first bracket in (\ref{e10}) be 
a Lorentz
condition $\partial_{0}A_{0}+\mbox{div}{\bf A}=0$. It is easily seen that the
other bracket be components of electric and magnetic fields:
$-E_{i}=-(\partial_{i}A_{0}+\partial_{0}A_{i}),\hspace{1mm}H_{i}=
(\mbox{curl\bf A})_{i}$.

Since $\omega=\e_{123}$ is belong to the center of $\R_{3}$, then
$$\omega \e_{1}=\e_{1}\omega=\e_{2}\e_{3},\hspace{2mm}\omega \e_{2}=\e_{2}
\omega=\e_{3}\e_{1},
\hspace{2mm}\omega \e_{3}=\e_{3}\omega=\e_{1}\e_{2}.$$

In accordance with these correlations may be written (\ref{e10}) as
\begin{equation}\label{e11}
{\cal A}_{0}{\cal A}_{1}=(E_{1}+\omega H_{1})\e_{1}+(E_{2}+\omega H_{2})
\e_{2}+(E_{3}+\omega H_{3})\e_{3}
\end{equation}

It is obvious that the expression (\ref{e11}) is coincide with the vector part
of complex quaternion (hyperbolic biquaternion) when $\e^{\p}_{1}=i\e_{1},
\e^{\p}_{2}=i\e_{2},\e^{\p}_{3}=i\e_{1}i\e_{2}$.

Further on, make up the exterior product $\bigtriangledown\mbox{\bf F}$, where
$\bigtriangledown$ is the first element from (\ref{e9}) and {\bf F} is an 
expression
of type (\ref{e10}):
$$\bigtriangledown\mbox{\bf F}=\mbox{div\bf E}\e_{0}-((\mbox{curl\bf H})_{1}-
\partial_{0}E_{1})\e_{1}-((\mbox{curl\bf H})_{2}-\partial_{0}E_{2})\e_{2}-$$
\begin{equation}\label{e12}
-((\mbox{curl\bf H})_{3}-\partial_{0}E_{3})\e_{3}+((\mbox{curl\bf E})_{1}+
\partial_{0}H_{1})\e_{2}\e_{3}+((\mbox{curl\bf E})_{2}+\partial_{0}H_{2})
\e_{3}\e_{1}+
\end{equation}
$$+((\mbox{curl\bf E})_{3}+\partial_{0}H_{3})\e_{1}\e_{2}+\mbox{div\bf H}
\e_{1}\e_{2}\e_{3}.$$

It is easily seen that the first coefficient of the product 
$\bigtriangledown\mbox{\bf F}$
be a left part of equation $\mbox{div\bf E}=\varrho$. The following three
coefficients are make up a left part of equation $\mbox{curl\bf H}-
\partial_{0}\mbox{\bf E}=j$,
the other coefficients are make up the equations 
$\mbox{curl\bf E}+\partial_{0}\mbox{\bf H}=0$
and $\mbox{div\bf H}=0$, respectively (for more details see \cite{7}).
\section{MODULO 2 PERIODICITY OF COMPLEX \protect\newline CLIFFORD ALGEBRAS}
In dependence from the sign of the square of volume element 
$\omega$ all totality
of Clifford algebras is divided into two classes. Namely, {\it if 
$\omega^{2}=1$ Clifford algebra ${}^{l}\K_{n}$ over a field $\K\hspace{2mm}
(\K=\R, \K=\Om, \K=\C)$ is called positive, and negative in contrary case
$(\omega^{2}=-1)$}. Further on, the each algebra ${}^{l}\K_{n}$ is associated to
a vector space $V$ over a field $\K$. This space is endowed with a 
nondegenerate
quadratic form $Q$. It is obvious that the dimensionality $\dim V$ of vector
space is equal to a number of units of the algebra ${}^{l}\K_{n}$. In 
accordance with this most general definition of Clifford algebra we shall
denote this algebra as ${}^{l}\K_{n}(V,Q)$.
\begin{theorem}[{Karoubi \cite{8}}] 1)\label{t1} 
If ${}^{l_{1}}\K_{n_{1}}(V_{1},Q_{1})$ is positive,
and if $\dim V_{1}$ is even, then
$${}^{l_{1}+l_{2}}\K_{n_{1}+n_{2}}(V_{1}\oplus V_{2},Q_{1}\oplus Q_{2})\cong
{}^{l_{1}}\K_{n_{1}}(V_{1},Q_{1})\otimes{}^{l_{2}}\K_{n_{2}}(V_{2},Q_{2}).$$
2) If ${}^{l_{1}}\K_{n_{1}}$ is negative, and if $\dim V_{1}$ is even, then
$${}^{l_{1}+l_{2}}\K_{n_{1}+n_{2}}(V_{1}\oplus V_{2},Q_{1}\oplus Q_{2})\cong
{}^{l_{1}}\K_{n_{1}}(V_{1},Q_{1})\otimes{}^{l_{2}}\K_{n_{2}}(V_{2},-Q_{2}).$$
\end{theorem}
\begin{theorem}\label{t2} The algebra ${}^{l+m}\K_{n+2m}$ is isomorphic to 
$\M_{2^{m}}({}^{l}\K_{n})$.\end{theorem}

Indeed, it is well-known that the algebra ${}^{l}\K_{n}$ with even 
dimensionality
$(n=2\nu)$ is isomorphic to a matrix algebra $\M_{2^{\nu}}(\K)$. Further on,
since ${}^{1}\K_{2}\hspace{2mm}({}^{1}\R_{2},\hspace{1mm}{}^{1}\Om_{2},
\hspace{1mm}
{}^{1}\C_{2})$ is positive, then in accordance with theorem \ref{t1} we obtain
$$
{}^{l+1}\K_{n+2}\cong{}^{l}\K_{n}\otimes{}^{1}\K_{2}\cong\left\{
\begin{array}{c}
{}^{l}\R_{n}\otimes{}^{1}\R_{2}\\
{}^{l}\Om_{n}\otimes{}^{1}\Om_{2}\\
{}^{l}\C_{n}\otimes{}^{1}\C_{2}
\end{array}\right\}\cong\left\{
\begin{array}{c}
{}^{l}\R_{n}\otimes\M_{2}(\R)\\
{}^{l}\Om_{n}\otimes\M_{2}(\Om)\\
{}^{l}\C_{n}\otimes\M_{2}(\C)
\end{array}\right\}\cong\left\{
\begin{array}{c}
\M_{2}({}^{l}\R_{n})\\
\M_{2}({}^{l}\Om_{n})\\
\M_{2}({}^{l}\C_{n})
\end{array}\right.$$
Therefore,
\begin{equation}\label{e13}
{}^{l+m}\K_{n+2m}\cong{}^{l}\K_{n}\otimes\M_{2}(\K)\otimes\ldots\otimes
\M_{2}(\K)\cong{}^{l}\K_{n}\otimes\M_{2^{m}}(\K)\cong\M_{2^{m}}({}^{l}\K_{n}).
\end{equation}
\begin{theorem}\label{t3} The algebra $\C_{n+2}$ is isomorphic to 
$\M_{2}(\C_{n})$.
\end{theorem}
It is obvious that in the case of a field $\K=\C$ we have an isomorphism
${}^{l}\C_{n}\cong\C_{n}$. Further on, in accordance with theorem \ref{t1}
we obtain $\C_{n+2}\cong\C_{n}\otimes\C_{2}$ (the algebra $\C_{2}$ is positive
if $\omega=i\e_{12}, \e^{2}_{1}=\e^{2}_{2}=1$ or if $\omega=\e_{12},
\e^{2}_{1}=1, \e^{2}_{2}=-1$). Since $\C_{2}\cong\M_{2}(\C)$, hence it follows
that
\begin{equation}\label{e14}
\C_{n+2}\cong\C_{n}\otimes\C_{2}\cong\C_{n}\otimes\M_{2}(\C)\cong
\M_{2}(\C_{n}).
\end{equation}

From adduced above theorems immediately follows that for any Clifford algebra
with even dimensionality $(n=2\nu)$ over a field $\C$ there exists a
decomposition :
\begin{equation}\label{e15}
\C_{2\nu}\cong\underbrace{\C_{2}\otimes\C_{2}\otimes\ldots\otimes
\C_{2}}_{\scriptstyle\nu \hspace{1mm}\mbox{\protect\small times}}.
\end{equation}  

In case of odd dimensionality ($n=2\nu+1$) the algebra $\C_{2\nu+1}$ is
decomposed to a direct sum of two subalgebras which are isomorphic to 
$\C_{2\nu}$.
This decomposition is realized by means of mutually orthogonal idempotents
$$\varepsilon=\frac{1}{2}(1+\omega),\hspace{2mm}\varepsilon^{\p}=\frac{1}{2}
(1-\omega),$$
where $\omega=\e_{12\ldots 2\nu+1}$ and $\omega^{2}=1$; or
$$\varepsilon=\frac{1}{2}(1+i\omega),\hspace{2mm}\varepsilon^{\p}=\frac{1}{2}
(1-i\omega),$$
where $\omega=\e_{12\ldots 2\nu+1}$ and $\omega^{2}=-1$. It is easily seen that
$\varepsilon + \varepsilon^{\p}=1,\hspace{1mm}\varepsilon\varepsilon^{\p}=0,
\varepsilon^{2}=\varepsilon,\hspace{1mm}{\varepsilon^{\p}}^{2}=
\varepsilon^{\p}$.
Since in this case volume element $\omega=\e_{12\ldots 2\nu+1}$ is belong to
a center of $\C_{2\nu+1}$, then the idempotents $\varepsilon,\hspace{1mm}
\varepsilon^{\p}$ are commutes with all elements of the algebra $\C_{2\nu+1}$.
Hence it follows that \cite{9}
$$\C_{2\nu+1}\cong\C_{2\nu}\oplus\C_{2\nu},$$
where the each algebra $\C_{2\nu}$ consist of the elements of type
$\varepsilon{\cal A}$ or $\varepsilon^{\p}{\cal A}$, here 
${\cal A}\in\C_{2\nu+1}$.

Therefore, in accordance with (\ref{e15}) we have
\begin{equation}\label{e16}
\C_{2\nu+1}\cong\C_{2}\otimes\C_{2}\otimes\ldots\otimes\C_{2}\oplus
\C_{2}\otimes\C_{2}\otimes\ldots\otimes\C_{2}.
\end{equation}

Further on, by force of $\C_{2}\cong\M_{2}(\C)$ we have for a base of spinor
representation of $\C_{2}$ the following matrices which are correspond to the
units $\e_{1}$ and $\e_{2}$:
\begin{equation}\label{e17}
E_{1}=\left[
\begin{array}{cc}
0 & 1 \\
1 & 0
\end{array}\right],\hspace{2mm}E_{2}=\left[
\begin{array}{cc}
0 & -i \\
i & 0
\end{array}\right].
\end{equation}
Let $\e_{3}=\e_{1}\e_{2}$, then for a vector part of biquaternion (\ref{e13})
we obtain by means of (\ref{e17}) the following matrix:
\begin{equation}\label{e18}
\left[
\begin{array}{cc}
iF_{3} & F_{1}-iF_{2} \\
F_{1}+iF_{2} & -iF_{3}
\end{array}\right],
\end{equation}
here $F_{i}=E_{i}+iH_{i}$, since in this case $\omega\equiv i$.

This way, {\it the algebra $\C_{2}$ with general element ${\cal A}=F_{0}\e_{0}+
F_{1}\e_{1}+F_{2}\e_{2}+F_{3}\e_{3}$ is isomorphic to a matrix algebra of
type (\ref{e18}), where $F_{0}=\partial_{0}A_{0}+\mbox{div\bf A}\equiv 0$ and
$F_{i}\hspace{1mm}(i=1,2,3)$ are components of a complex electromagnetic 
field.}
We shall denote this matrix algebra as $\M^{F}_{2}(\C)$.

Hence by force of modulo 2 periodicity of complex Clifford algebras from
(\ref{e15}) and (\ref{e16}) immediately follows that
\begin{eqnarray}
\C_{2\nu}&\cong&\M^{F}_{2^{\nu}}(\C)\cong\M^{F}_{2}(\C)\otimes\M_{2}^{F}(\C)
\otimes\ldots\otimes\M^{F}_{2}(\C),\label{e19} \\
\hspace{2mm}\C_{2\nu+1} & \cong & \C_{2\nu}\oplus\C_{2\nu}
\cong\M_{2^{\nu}}^{F}(\C)\oplus
\M_{2^{\nu}}^{F}(\C)\cong \label{e20} \\
&\cong & \M^{F}_{2}(\C)\otimes\M_{2}^{F}(\C)\otimes\ldots\otimes\M^{F}_{2}(\C)
\oplus\M^{F}_{2}(\C)\otimes\M^{F}_{2}(\C)\otimes\ldots\otimes\M^{F}_{2}(\C).
\nonumber
\end{eqnarray}

For example, consider now the algebra $\C_{4}$. In the spinor representation
this algebra is isomorphic to a matrix algebra $\M_{4}(\C)$, the base of
which consist of well-known $\gamma$-matrices. In the base of Weyl for these
matrices we have
\begin{equation}\label{e21}
\gamma^{m}=\left[
\begin{array}{cc}
0 & \sigma^{m} \\
\overline{\sigma}^{m} & 0
\end{array}\right],\end{equation}
where $m=0,1,2,3$ and
$$\sigma^{0}=\left[
\begin{array}{cc}
1 & 0 \\
0 & 1
\end{array}\right],\hspace{1mm}\sigma^{1}=\left[
\begin{array}{cc}
0 & 1 \\
1 & 0 
\end{array}\right],\hspace{1mm}\sigma^{2}=\left[
\begin{array}{cc}
0 & -i \\
i & 0
\end{array}\right],\hspace{1mm}\sigma^{3}=\left[
\begin{array}{cc}
1 & 0 \\
0 & -1
\end{array}\right],$$
$$\overline{\sigma}^{0}=\sigma^{0},\hspace{1mm}\overline{\sigma}^{1,2,3}=
-\sigma^{1,2,3}.$$
It is easily seen that $\sigma$-matrices are make up the base of spinor
representation of $\C_{2}\cong\M_{2}(\C)$ and are coincide with matrices
$E_{i}$ if $\sigma^{3}=\frac{1}{i}E_{1}E_{2}$.

For the algebra $\C_{4}$ in accordance with (\ref{e19}) we have
$$\C_{4}\cong\C_{2}\otimes\C_{2}\cong\M^{F}_{2}(\C)\otimes\M_{2}^{F}(\C).$$
Further on, by general definition, the algebra $\C_{4}$ is associated to a
complex vector space $C_{4}$. For the vector $\mbox{\bf a}=F_{i}\e_{i}\in 
C_{4}$
in the base (\ref{e21}) we have a matrix $\mbox{\bf A}=F_{i}\gamma^{i}$:
$$\left[
\begin{array}{cccc}
0 & 0 & F_{3} & F_{1}-iF_{2} \\
0 & 0 & F_{1}+iF_{2} & -F_{3} \\
-F_{3} & -F_{1}+iF_{2} & 0 & 0 \\
-F_{1}-iF_{2} & F_{3} & 0 & 0
\end{array}\right].$$
The all matrix algebra $\M_{4}^{F}(\C)$ in the base (\ref{e17}) we obtain 
by means of theorem \ref{t3}.
Indeed, in the case when dimensionality of ${}^{l}\K_{n}$ is even the volume
element $\omega=\e_{12\ldots n}$ is not belong to a center of algebra
${}^{l}\K_{n}$.

However, when $i\leq 2m$ we have
\begin{eqnarray}
\e_{12\ldots 2m 2m+k}\e_{i}&=&(-1)^{2m+1-i}\sigma(i-l)
\e_{12\ldots i-1 i+1\ldots 2m 2m+k},\nonumber \\
\e_{i}\e_{12\ldots 2m 2m+k}&=&(-1)^{i-1}\sigma(i-l)
\e_{12\ldots i-1 i+1\ldots 2m 2m+k}\nonumber
\end{eqnarray}
and, therefore, the condition of commutativity of elements 
$\e_{12\ldots 2m 2m+k}$
and $\e_{i}$ is $2m+1-i\equiv i-1\!\!\!\!\pmod{2}$. Thus, the elements 
$\e_{12\ldots 2m 2m+1}$ and $\e_{12\ldots 2m 2m+2}$ are commute with all
elements $\e_{i}$ whose indexes are not exceed $2m$. Therefore, the transition
from algebra $\K_{2m}$ to algebras $\K_{2m+2},\;{}^{1}\K_{2m+2}$ or 
${}^{2}\K_{2m+2}$ and from algebra ${}^{l}\K_{2m}$ to ${}^{l}\K_{2m+2},\;
{}^{l+1}\K_{2m+2}$ or ${}^{l+2}\K_{2m+2}$ may be represented as transition
from the real (complex or double) coordinates of elements of algebras 
$\K_{2m}$ and ${}^{l}\K_{2m}$
to the coordinates of type $a+b\phi+c\psi+d\phi\psi$, where $\phi$ and $\psi$
are additional basis elements $\e_{12\ldots 2m 2m+1}$ and 
$\e_{12\ldots 2m 2m+2}$. The elements $\e_{i_{1}i_{2}\ldots i_{k}}\phi$
are contain index $2m+1$ and not contain index $2m+2$, and the elements
$\e_{i_{1}i_{2}\ldots i_{k}}\psi$ are contain index $2m+2$ and not contain
index $2m+1$. Respectively, the elements $\e_{i_{1}i_{2}\ldots i_{k}}
\phi\psi$ are contain both indexes $2m+1$ and $2m+2$. Hence it immediately
follows that a general element of ${}^{l}\K_{2m+2}$ or ${}^{l+1}\K_{2m+2},\;
{}^{l+2}\K_{2m+2},\;\K_{2m+2}$ may be represented as
\begin{equation}\label{e21'}
{}^{l}\K^{0}_{2m}\e_{0}+{}^{l}\K^{1}_{2m}\e_{12\ldots 2m 2m+1}+
{}^{l}\K^{2}_{2m}\e_{12\ldots 2m 2m+2}+{}^{l}\K^{3}_{2m}\e_{2m+1 2m+2},
\end{equation}
where ${}^{l}\K^{i}_{2m}\;(i=0,1,2,3)$ are represent the algebras with
general element ${\cal A}=\sum^{2m}_{k=0}a^{i_{1}i_{2}\ldots i_{k}}
\e_{i_{1}i_{2}\ldots i_{k}}$. When the elements $\phi=\e_{12\ldots 2m 2m+1}$
and $\psi=\e_{12\ldots 2m 2m+2}$ are satisfy to condition 
$\phi^{2}=\psi^{2}=-1$ we see that the basis $\{\e_{0},\,\phi,\,\psi,\,
\phi\psi\}$ is isomorphic to a basis of quaternion algebra. Therefore,
in this case from (\ref{e21'}) we have
$${}^{l}\K_{2m+2}\cong\M_{2}({}^{l}\K_{2m})=\left[
{\renewcommand{\arraystretch}{1.3}
\begin{array}{cc}
{}^{l}\K^{0}_{2m}-i{}^{l}\K^{3}_{2m} & -{}^{l}\K^{1}_{2m}+i{}^{l}\K^{2}_{2m} \\
{}^{l}\K^{1}_{2m}+i{}^{l}\K^{2}_{2m} &\phantom{-}{}^{l}\K^{0}_{2m}+
i{}^{l}\K^{3}_{2m}
\end{array}}\right].$$
Analoguosly, when the elements $\phi$ and $\psi$ are satisfy to conditions
$\phi^{2}=\psi^{2}=1$ or $\phi^{2}=1,\;\psi^{2}=-1$ we have the transitions
from the real (complex or double) coordinates in ${}^{l}\K_{2m+2}$ to the
anti-quaternionic and pseudo-quaternionic coordinates, respectively
(about anti-quaternions and pseudo-quaternions see \cite[p. 434]{10}).  
\begin{sloppypar}
In our case by force of isomorphism ${}^{l}\C_{n}\cong\C_{n}$ we see that
any algebra ${}^{l}\C_{4}\cong\C_{4}\;(l=0,1,2,3,4)$ may be represented
by the following quaternion:
$$\C^{0}_{2}\e_{0}+\C^{1}_{2}\phi +\C^{2}_{2}\psi +\C^{3}_{2}\phi\psi,$$
where $\phi=\e_{123},\;\psi=\e_{124}$ and $\C^{i}_{2}$ the algebras of
hyperbolic biquaternions with general element (in our case)
${\cal A}=F^{0}\e_{0}+F^{1}\e_{1}+F^{2}\e_{2}+F^{3}\e_{12}$. In spinor
representation for the quaternion basis $\{\e_{0},\,\phi,\,\psi,\phi\psi\}$
we have $\left\{\left[\begin{array}{cc} 1 & 0 \\ 0 & 1 \end{array}\right],\;
\left[\begin{array}{cc} 0 & -1 \\ 1 & 0 \end{array}\right],\;
\left[\begin{array}{cc} 0 & i \\ i & 0 \end{array}\right],\;
\left[\begin{array}{cc} -i & 0 \\ 0 & i \end{array}\right]\right\}$.
Hence it immediately follows that\end{sloppypar}
$$\C_{4}\cong\C_{2}\otimes\C_{2}\cong\C_{2}\otimes\M_{2}(\C)\cong
\M_{2}(\C_{2})=\left[
{\renewcommand{\arraystretch}{1.3}
\begin{array}{cc}
\C^{0}_{2}-i\C^{3}_{2} & -\C^{1}_{2}+i\C^{2}_{2} \\
\C^{1}_{2}+i\C^{2}_{2} & \phantom{-}\C^{0}_{2}+i\C^{3}_{2}
\end{array}}\right].$$
Finally, using (\ref{e18}) we obtain that $\M^{F}_{4}(\C)$ has a form:
\begin{equation}\label{e23}
\left[
\begin{array}{cccc}\scriptstyle
(1+i)F_{3} & \s(1-i)F_{1}-(1+i)F_{2} & \s-(1+i)F_{3} & \s-(1-i)F_{1}+
(1+i)F_{2} \\
\s(1-i)F_{1}+(1+i)F_{2} & \s-(1+i)F_{3} & \s-(1-i)F_{1}-(1+i)F_{2} & 
\s(1+i)F_{3} \\
\s-(1-i)F_{3} & \s(1+i)F_{1}+(1-i)F_{2} & \s-(1-i)F_{3} & \s(1+i)F_{1}+
(1-i)F_{2} \\
\s(1+i)F_{1}-(1-i)F_{2} & \s(1-i)F_{3} & \s(1+i)F_{1}-(1-i)F_{2} & 
\s(1+i)F_{3} 
\end{array}\right]\end{equation}

Analoguosly, for the algebra $\C_{6}\cong\M_{8}^{F}(\C)$ we have the
anti-quaternion $\C^{0}_{4}\e_{0}+\C^{1}_{4}\phi+\C^{2}_{4}\psi
+\C^{3}_{4}\phi\psi$, where $\phi=\e_{12345},\;\psi=\e_{12346}$ and
$\phi^{2}=\psi^{2}=1$. The anti-quaternion basis $\{\e_{0},\;\phi,\;\psi,\;
\phi\psi\}$ in spinor representation defined by the following matricies
$\left\{\left[\begin{array}{cc} 1 & 0 \\ 0 & 1 \end{array}\right],\;
\left[\begin{array}{cc} 0 & 1 \\ 1 & 0 \end{array}\right],\;
\left[\begin{array}{cc} 0 & -i \\ i & 0 \end{array}\right],\;
\left[\begin{array}{cc} i & 0 \\ 0 & -i \end{array}\right]\right\}$.
Using (\ref{e23}) we have
$$2\left[
\begin{array}{cccccccc}
\s iF_{3} & \s F_{1}-iF_{2} & \s-iF_{3} & \s-F_{1}+iF_{2} & \s F_{3} & 
\s-F_{2}-iF_{1} & \s-F_{3} & \s F_{2}+iF_{1} \\
\s F_{1}+iF_{2} & \s-iF_{3} & \s-F_{1}-iF_{2} & \s iF_{3} & \s F_{2}-iF_{1} & 
\s-F_{3} & \s F_{2}+iF_{1} & \s F_{3}  \\
\s-F_{3} & \s F_{2}+iF_{1} & \s-F_{3} & \s F_{2}+iF_{1} & \s iF_{3} & 
\s F_{1}-iF_{2} & \s iF_{3} & \s F_{1}-iF_{2} \\
\s-F_{1}+iF_{1} & \s F_{3} & \s-F_{2}+iF_{1} & \s F_{3} & \s F_{1}+iF_{2} & 
\s-iF_{3} & \s F_{1}+iF_{2} & \s-iF_{3} \\
\s iF_{3} & \s F_{1}-iF_{2} & \s-iF_{3} & \s-F_{1}+iF_{2} & \s F_{3} & 
\s-F_{2}-iF_{1} & \s-F_{3} & \s F_{2}+iF_{1} \\
\s F_{1}+iF_{2} & \s-iF_{3} & \s-F_{1}-iF_{2} & \s iF_{3} & \s F_{2}-iF_{1} & 
\s-F_{3} & \s F_{2}+iF_{1} & \s F_{3} \\
\s-F_{3} & \s F_{2}+iF_{1} & \s-F_{3} & \s F_{2}+iF_{1} & \s iF_{3} & 
\s F_{1}-iF_{2} & \s iF_{3} & \s F_{1}-iF_{2} \\
\s-F_{1}+iF_{1} & \s F_{3} & \s-F_{2}+iF_{1} & \s F_{3} & \s F_{1}+iF_{2} & 
\s-iF_{3} & \s F_{1}+iF_{2} & \s-iF_{3}
\end{array}\right].$$
For the algebras with odd dimensionality from (\ref{e20}) we obtain
$\C_{5}\cong\M^{F}_{4}(\C)\oplus\M^{F}_{4}(\C)$, $\C_{7}\cong\M_{8}^{F}(\C)
\oplus\M^{F}_{8}(\C)$ and so on.

\end{document}